\documentclass[10pt, a4paper]{article}

\usepackage{header}

\usepackage{lrec2022} 
\usepackage{multibib}
\newcites{languageresource}{Language Resources}
\usepackage{graphicx}
\usepackage{tabularx}
\usepackage{soul}
\usepackage{titlesec}
\titleformat{\section}{\normalfont\large\bfseries\center}{\thesection.}{1em}{}
\titleformat{\subsection}{\normalfont\SmallTitleFont\bfseries\raggedright}{\thesubsection.}{1em}{}
\titleformat{\subsubsection}{\normalfont\normalsize\bfseries\raggedright}{\thesubsubsection.}{1em}{}
\renewcommand\thesection{\arabic{section}}
\renewcommand\thesubsection{\thesection.\arabic{subsection}}
\renewcommand\thesubsubsection{\thesubsection.\arabic{subsubsection}}

\usepackage{epstopdf}
\usepackage[utf8]{inputenc}
\usepackage{hyperref}
\usepackage{xstring}
\usepackage{color}
\usepackage{cleveref}
\usepackage{algpseudocode}
\usepackage{booktabs}
\usepackage{xspace}
\usepackage{float}
\usepackage{graphicx}
\usepackage{subcaption}
\usepackage{arydshln}

\newcommand{\projectName}{D3\xspace}
\newcommand{\nlpScholar}{NLP Scholar\xspace}
\newcommand{\nlpExplorer}{NLPExplorer\xspace}
\newcommand{\googleScholar}{Google Scholar\xspace}
\newcommand{\semanticScholar}{Semantic Scholar\xspace}
\newcommand{\aclAnthology}{ACL Anthology\xspace}
\newcommand{\drift}{DRIFT\xspace}
\newcommand{\nlpfournlp}{NLP4NLP\xspace}

\definecolor{darkblue}{rgb}{0, 0, 0.5}
\hypersetup{colorlinks=true, citecolor=darkblue, linkcolor=darkblue, urlcolor=darkblue}

\title{\projectName: A Massive Dataset of Scholarly Metadata\\ for Analyzing the State of Computer Science Research}

\name{Jan Philip Wahle\textsuperscript{\textdagger}, Terry Ruas\textsuperscript{\textdagger}, Saif M. Mohammad\textsuperscript{\textdagger\textdagger}, Bela Gipp\textsuperscript{\textdagger}} 

\address{\textsuperscript{\textdagger}University of Wuppertal Germany, \textsuperscript{\textdagger\textdagger}National Research Council Canada \\
        \textsuperscript{\textdagger}\{wahle, ruas, gipp\}@uni-wuppertal.de\\
        \textsuperscript{\textdagger\textdagger}saif.mohammad@nrc-cnrc.gc.ca\\
}

\abstract{
DBLP is the largest open-access repository of scientific articles on computer science and provides metadata associated with publications, authors, and venues.
We retrieved more than 6 million publications from DBLP and extracted pertinent metadata (e.g., abstracts, author affiliations, citations) from the publication texts to create the DBLP Discovery Dataset (\projectName).
\projectName can be used to identify trends in research activity, productivity, focus, bias, accessibility, and impact of computer science research.
We present an initial analysis focused on the volume of computer science research (e.g., number of papers, authors, research activity), trends in topics of interest, and citation patterns.
Our findings show that computer science is a growing research field ($\approx$15\% annually), with an active and collaborative research community.
While papers in recent years present more bibliographical entries in comparison to previous decades, the average number of citations has been declining.
Investigating papers' abstracts reveals that recent topic trends are clearly reflected in \projectName.
Finally, we list further applications of \projectName and pose supplemental research questions.
The \projectName dataset, our findings, and source code are publicly available for research purposes.
\\ \newline  }

\begin{document}

\maketitleabstract

\thispagestyle{firststyle}

\section{Introduction}
In the last few decades, computer science (CS) has transformed many scientific fields.
Faster systems, more accurate results, and efficient tools are just some of the benefits provided by computer science advancements.
Arguably, today there is hardly any area not affected by its vast possibilities.
Consider how difficult it would be to test, develop, and research new vaccines without access to tools of informatics (e.g., public repositories).

The techniques behind these contributions are often made available through scientific publications which can be used to investigate trends and patterns within computer science itself.
However, computer science is a large field with many sub-areas (e.g., natural language processing (NLP), computer vision) and repositories (e.g., arXiv); thus, a complete analysis of its publications is a challenging task.
How fast is computer science research growing? How many authors are actively publishing in their field? What topics are prevalent in specific venues? 
A large and carefully curated dataset of CS-publications metadata is crucial for the quantitative exploration of these questions.

DBLP is one of the largest open computer science repositories with records from major journals and proceedings starting from 1936.\footnote{\url{https://dblp.org/}}
The repository provides access to several pieces of metadata associated with each of its papers, including author names, title, year, and venue. 
The papers stored in DBLP come from paid publishers (e.g., IEEE, ACM) and open repositories (e.g., ACL, arXiv).
\nlpScholar\cite{Mohammad20b} and arXiv also offer an extensive collection of papers in computer science, but both are included in DBLP.
Therefore DBLP offers a more complete corpus to understand patterns in computer science.
However, as DBLP mainly indexes metadata of papers, it lacks some important information that is embedded in their full texts (e.g., affiliations, citations).

In this paper, we introduce the DBLP Discovery Dataset (\projectName), which not only includes key information about papers from DBLP in an easily accessible form, but it also enriches it with crucial metadata such as abstracts, author affiliations, and citations, that we extracted from the full texts. Thus, \projectName can be used to explore and understand broad trends in computer science research.

Our dataset is proposed as a lightweight resource to explore the patterns in computer science publications and the relation between the elements describing them, e.g., what are the most popular topics of conferences in 2021? How active have authors been over the years, and how long, on average, are authors active?
\projectName can also be used as a training corpus in language modeling, classification of papers by topic and venues, statistical analysis of publishers, and several other scenarios.
Even though \googleScholar and \semanticScholar are similar to \projectName, they only provide access to their metadata. 
For example, while one can access the number of citations and information on which paper cites which paper individually, one cannot access the data in bulk for large-scale quantitative analysis.
In addition, the access to the actual dataset in \googleScholar and \semanticScholar is not straightforward.
For example, \googleScholar has no standard API and limits its web page access to crawl.
Although the Semantic Scholar Open Research Corpus (S2ORC) offers a dump from 2020 \cite{lo-wang-2020-s2orc}, its 81 million papers require more than 800GB of storage, which can be restrictive to many researchers trying to process and analyze the data.
In comparison, the uncompressed size of \projectName is $\approx 18$GB.

In summary, our contributions are two-fold. 
We (1) publish a new open dataset\footnote{\url{https://doi.org/10.5281/zenodo.7069915}} of $\approx 6$ million English research papers and the source code to retrieve them\footnote{\url{https://bit.ly/d3-dataset}}, and (2) provide an initial investigation of computer science publications.
\projectName augments DBLP with metadata extracted from full-text to provide additional features over existing datasets such as paper abstracts and author affiliations (see \Cref{tab:secondary_full_text_attributes} for more details).
We provide an exploratory analysis of \projectName through eight research questions to illustrate some of our dataset's main capabilities.
Our questions investigate the volume, content, and citations of papers in \projectName.

\section{Existing Resources}
The \nlpScholar dataset \cite{Mohammad20b} combines primary information from $\approx45,000$ NLP publications in the \aclAnthology (e.g., authors, venues) with citation information from \googleScholar.
Mohammad \shortcite{Mohammad20b,Mohammad20d} used the \nlpScholar dataset to examine citation patterns, the gender gap between female and male first-time authors, and $n-$gram distributions through interactive visualization. 
\drift \cite{SharmaCPP21} tracks the changes in \textit{cs.CL}, the computer science computational linguistics tag from arXiv, focusing on single-word terms and their word embeddings over time.
In \nlpExplorer, \cite{ParmarJJJ20} provide an exploratory tool for NLP publications based on \aclAnthology, including information on most-cited authors, areas, and venues; similar to \nlpScholar. As of June 2021, \nlpExplorer also explores Tweets related to publications and conferences.
The \nlpfournlp Corpus \cite{Mariani_Francopoulo_Paroubek_2019} contains $\approx65,000$ articles from 34 conferences and journals in NLP such as the \aclAnthology and the International Speech Communication Association.
They provide extensive analysis on references and volume, citations, and authorship.
S2ORC \cite{lo-wang-2020-s2orc} is a repository of 81.1 million English academic papers from 20 research fields like medicine, biology, or physics.
Apart from citations, semantic scholar also offers information about venues and authors.

We extend current datasets, i.e., \nlpScholar, \drift, in two key directions. We (1) include computer science venues outside of the \aclAnthology and arXiv, and (2) add informative features derived from full-texts (e.g., citations, paper abstracts, or author affiliations).
As NLP research is not restricted to a single repository, \projectName provides a more comprehensive view on the entirety of NLP research and many other subfields in computer science.
Our dataset contains records from many publishers (e.g., Springer, IEEE, ACM), including the entire \aclAnthology and computer science publications on arXiv, thus allowing \projectName to answer all questions from previous datasets more accurately.

\section{Dataset Collection}
We extracted all records from DBLP and crawled their associated full-text PDFs to extract metadata (e.g., bibliographies) from January 1st, 1936 until December 2nd, 2021.\footnote{DBLP is monthly updated with new records.}
The subsections below describe how we extracted and aligned DBLP and full-text information.

\subsection{Primary Information from DBLP}

DBLP provides open access to its data in two ways, a public search API\footnote{\url{https://dblp.org/search/publ/api}}, and a XML dump\footnote{\url{https://dblp.org/xml/release}}.
We are interested in understanding the state of computer science research at a large scale over time, so we retrieve their full XML release.
To keep \projectName up-to-date, we download the latest DBLP release monthly and compute the changes to our last version of \projectName.
We provide an overview of the attributes with examples that we retrieve from DBLP in \Cref{tab:primary_dblp_attributes}.

\begin{table}[tb]
\small
\centering
\begin{tabular}{lr} 
\toprule
\textbf{Attribute} &\textbf{Example}\\
\toprule
publication                          & \\
\hspace{3mm} id                      & conf/acl/Mohammad20b\\
\hspace{3mm} modified date           & 2021-09-12\\
\hspace{3mm} title                   & NLP Scholar - An Interactive ...\\
\hspace{3mm} pages                   & 232-255\\
\hspace{3mm} year                    & 2020\\
\hspace{3mm} type                    & Conference and Workshop Papers\\
\hspace{3mm} access                  & open\\
\hspace{3mm} links                   & [https://doi.org/...]\\
\hspace{3mm} doi                     & 10.18653/v1/2020.acl-demos.27\\
\hspace{3mm} publisher               & ACL\\
author                              & \\
\hspace{3mm} id                      & 58/380\\
\hspace{3mm} fullname                & Saif M. Mohammad\\
\hspace{3mm} webpage                 & http://saifmohammad.com/\\
venue & \\
\hspace{3mm} names                   & [International Conference on Lang...] \\
\hspace{3mm} acronyms                & [LREC] \\
\hspace{3mm} type                    & Conference or Workshop \\
\hspace{3mm} id                     & conf/lrec \\
\bottomrule
\end{tabular}
\caption{Primary attributes of \projectName.} \label{tab:primary_dblp_attributes}
\end{table}


\textbf{Publications.} The majority of entries in DBLP are indexed publications with metadata.
Examples for other entries are web pages and author information.
The dataset classifies papers according to the BibTeX entry types\footnote{\url{https://bit.ly/33vqaVG}} (e.g., article, in proceedings).
We transform publications by paper type into a standard JSON format and map authors and venues to uniquely identified entities.\\[1.5pt]
\textbf{Authors.} DBLP processes multiple authors with the same name using an iterative counter and the same authors with various variants of their names with unified entities.
We create a unique id for each author to map them to their publications.
Author entities in DBLP are sparse and typically only provide a personal webpage URL but rarely other informative features such as a current affiliation.
To enhance authors' entities with beneficial features, we extract information from full-texts (e.g., affiliations) in \Cref{sec:secondary_information}.\\[1.5pt]
\textbf{Venues.} For most publications, DBLP provides a venue code (i.e., the abbreviation of the venue).
We create venue entities for each and map them to papers by generating a unique id.
DBLP also contains information about major publishers such as Springer, IEEE, and ACM.
As the data on publishers is scarce ($\approx$10\% or publications have publishers annotated), we store publishers directly in the publication entries.

\begin{figure*}[!ht]
    \centering
    \includegraphics[width=\textwidth]{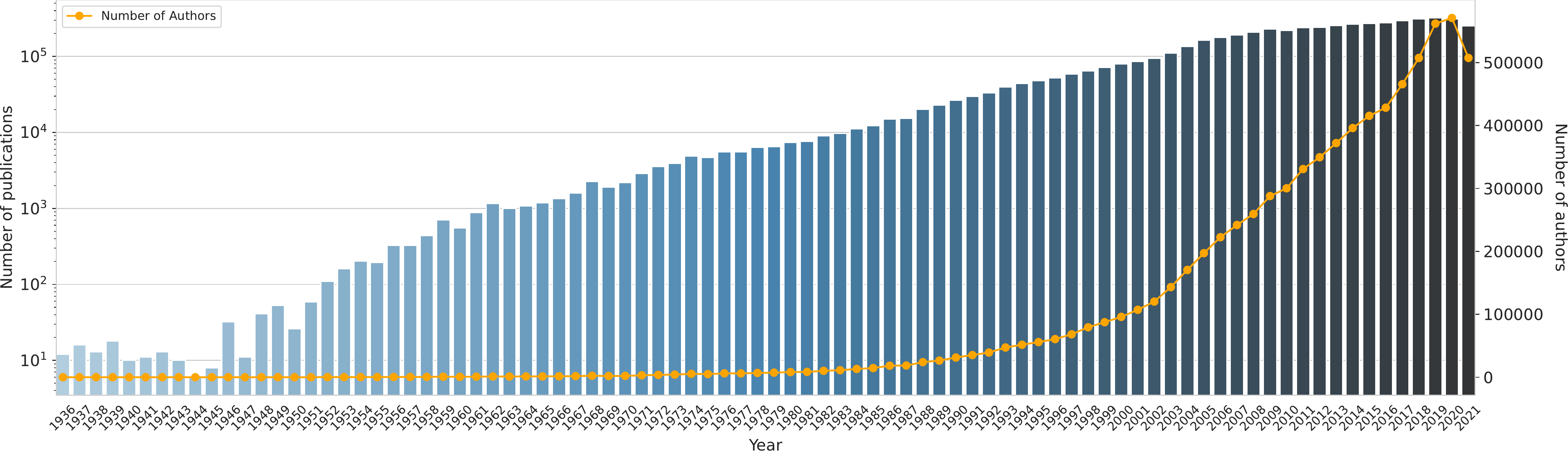}
    \caption{The number of annual publications and authors in logarithmic scale between 1936 and 2021.}
    \label{fig:analysis-entries-over-years}
\end{figure*}

\subsection{Secondary Information from Full-Texts} \label{sec:secondary_information}
\begin{table}[tb]
\small
\centering
\begin{tabular}{lr} 
\toprule
\textbf{Attribute} & \textbf{Example}\\
\toprule
affiliations                & \\
\hspace{3mm} id             & 4eb3...f094\\
\hspace{3mm} name           & National Research Council Canada\\
\hspace{3mm} country        & Canada\\
\hspace{3mm} city           & Ottawa\\
\hspace{3mm} postcode       & K1A 0R6\\
\hspace{3mm} addressline    & 1200 Montreal Road, Bldg. M-58\\
outgoing citations          & \\
\hspace{3mm} ids            & [7615..., 76af...]\\
\hspace{3mm} count          & 2\\
incoming citations           & \\
\hspace{3mm} ids            & [7ca5..., 7d0e...] \\
\hspace{3mm} count          & 11\\
keywords                    & [Scientometrics, Citations, ...]\\
ocr title                       & NLP Scholar: An Interactive ...\\
ocr abstract                    & As part of the NLP Scholar ...\\

\bottomrule
\end{tabular}
\caption{Secondary attributes of \projectName.} \label{tab:secondary_full_text_attributes}
\end{table}
Publications' full-texts contain valuable information about author affiliations, content, and references not present in DBLP, other datasets (e.g., \nlpScholar), or online services (e.g., \googleScholar).
We provide an overview of the attributes with example values that we extract from full-texts in \Cref{tab:secondary_full_text_attributes}.\\[1.5pt]
\textbf{Abstracts.} We retrieve the abstracts and index keywords of publications using GROBID \cite{GROBID}. For abstract extraction, the model uses CRF Wapiti \cite{lavergne2010practical} features and achieves an F1-score (using Levenshtein Matching with minimum distance of 0.8) of 92.85\% when drawing 1943 PubMed\footnote{\url{https://pubmed.ncbi.nlm.nih.gov/}} papers\footnote{\url{https://bit.ly/3K9Vf2g}}.
Using this model, we retrieved 4,219,855 abstracts from papers which are 78.17\% of the dataset.
The remaining papers were either disregarded by GROBID because of poor quality or did not provide an accessible document that could have been parsed. 
We use the publication's unique id to align the extracted information with DBLP. \\[1.5pt]
\textbf{Affiliations.} We extract the author names and affiliations with the same model and sequence features as used for extracting the abstracts.
To create author--affiliation pairs, we match author names from extracted affiliations to author names in DBLP using Levenshtein distance.
In practice, creating author--affiliation pairs through name matching is robust (we found less than 5\% cases where it fails).
We demonstrate this by performing two small bootstrap and permutation tests \cite{dror-etal-2018-hitchhikers}.
In the first test, we draw 20 samples of $n=100$ publications uniformly at random and evaluate how often author names extracted from the PDFs do not match those in DBLP.
To draw more challenging samples in the second test, we took the first $n=100$ publications from a ranked list in which the average Levenshtein distance between authors' names was increasing.
Both tests show less than 5\% of names are mismatched ($p<0.001$).\\[1.5pt]
\textbf{Citations.} To collect the citations within DBLP, we build a citation graph from the bibliographies in full-texts similar to the Reference Corpus of the \aclAnthology \cite{radev-etal-2009-acl}.
To parse the publications' bibliographies, we use GROBID's BidLSTM-CRF features, which obtain an F1-score of 87.73\% for the PubMed samples (using Levenshtein Matching with minimun distance of 0.8)\footnote{\url{https://bit.ly/3K9Vf2g}}.
We build the citation links by adding two fields to each publication object, the incoming citations (i.e., how often a paper was cited) and outgoing citations (i.e., the number of bibliography entries in a paper).
The resulting citation graph allows us to investigate the role of authors, venues, publishers, and institutions in research. 
Additionally, their interaction will also help us identify implicit trends, common topics, and influence between the participants of this graph.
Even though \googleScholar offers an open access service, its data is restricted, preventing researchers from obtaining large-scale access to their citation information.
\googleScholar also does not have a standardized API and limits its access for crawling their webpage.
Linking citation within DBLP provides us with a focused view on the influence of sub-fields within computer science.
In \Cref{sec:analysis-citations}, we measure citations coming from fields other than computer science using \semanticScholar with the result that 21.15\% of citations are cited from papers outside of our repository (i.e., from other fields than computer science).

\subsection{Implementation Details} \label{sec:implementation}
Parsing releases of $\approx6$ million publications and extracting their metadata from full-texts is a resource-intensive process.
Therefore, we implemented a parallel and asynchronous routine to parse releases, retrieve full-texts, extract their metadata, and align the information to DBLP with a low disk, memory, and computational requirements.
First, we split the dataset into equally sized chunks as it allows us to work on mutually exclusive parts of the dataset with multiple processes at different times without processing the whole repository.
We launch $n$ processes to retrieve publications for $n$ chunks. Each process asynchronously requests the PDF link of a paper or, if not present, parses the HTML page of the paper to identify the PDF link.
Next, we download the PDF to a folder with its unique key.
As all requests are sent asynchronously, we reduce their idle times in between.
To restrict requests to the same domain and respect server limits, we use semaphores and wait to respect the retry-after header whenever we receive an HTTP 429 response.
In parallel to the $n$ retrieval processes, we launch another $n$ processes to work on full-texts from the previously downloaded chunk and extract their metadata.
To reduce disk requirements, we delete the full-texts after extraction and only keep their metadata.
The uncompressed size of \projectName, in JSON format, is $\approx 18$GB.

\section{Dataset Analysis}

\subsection{Volume \& Research Activity}

\textit{Q1. How large is DBLP? How does the number of publications change over time?}\\[1.5pt]
A. As of December 2021, DBLP contains a total of 6,392,734 entries.
\Cref{tab:analysis-entries-by-type} shows the number of publications by type until December 2021.
Most DBLP publications are either conference/workshop papers (47.12\%) or journal articles (43.42\%).
The third-largest category (6.05\%) is what DBLP refers to as "informal publications".
These are papers published in online repositories (such as arXiv) without a systematic peer review, as well as contributions to informal workshops.
A majority of these are arXiv pre-prints from the computing research repository (CoRR).
When informal publications are published in a peer-reviewed venue, DBLP updates its type accordingly.

\begin{table}[tb]
\small
\centering
\begin{tabular}{lrr} 
\toprule
\textbf{Paper Type} &\textbf{Count} & \textbf{Proportion}\\
\toprule
in proceedings          & 3,012,358     & 47.12\%\\
article                 & 2,776,011     & 43.42\%\\
informal                & 386,574       & 6.05\%\\
phd thesis              & 81,954        & 1.28\%\\
in collection           & 67,502        & 1.05\%\\
proceedings$^{\dagger}$             & 49,265        & 0.77\%\\
book                    & 19,070        & 0.30\%\\
\midrule
\textbf{total}          & \textbf{6,392,734}     & \textbf{100\%}\\
\bottomrule
\end{tabular}
\caption{The number of publications in DBLP by type until December 2021. \textsuperscript{\textdagger}The entire collection published.} \label{tab:analysis-entries-by-type}
\end{table}

DBLP contains 99.3\% of papers from the \aclAnthology.
The papers in \aclAnthology are concentrated between conferences and workshops (\textgreater90\%) \cite{Mohammad20b}, while DBLP provides a more balanced distribution between journal articles, conference, and workshop papers.
As DBLP contains the \aclAnthology and other non-ACL venues (e.g., IEEE, ACM), \projectName is a robust resource that enables the correlation between publications, venues, and authors in NLP and other areas in computer science.

The number of publications in computer science is growing on average 15.12\% yearly.
Between 1936 and 1952, the annual number of publications never exceeded one hundred papers as \Cref{fig:analysis-entries-over-years} shows.
However, after 1952 the number of publications start growing exponentially\footnote{The seeming drop in publications and authors in 2021 results from the dataset being crawled on December 2nd and therefore does not include the full month of December.}.\\

\textit{Q2. How many authors publish in \projectName over time? How has the average number of authors per paper changed over time?}\\[1.5pt]
A. There are  $\approx$2,9 million authors for the $\approx$6 million papers in \projectName.
The line in \Cref{fig:analysis-entries-over-years} shows the number of authors yearly between 1936 and 2021.
The number of authors in research papers increased significantly in 1990, showing that computer science became a popular research field.
Considering the advances in software and hardware in the last decades with the increasing dependency between computer science and other research areas, we expect the number of papers and authors to continue to grow in the following years.

\begin{figure}
    \centering
    \includegraphics[width=\columnwidth]{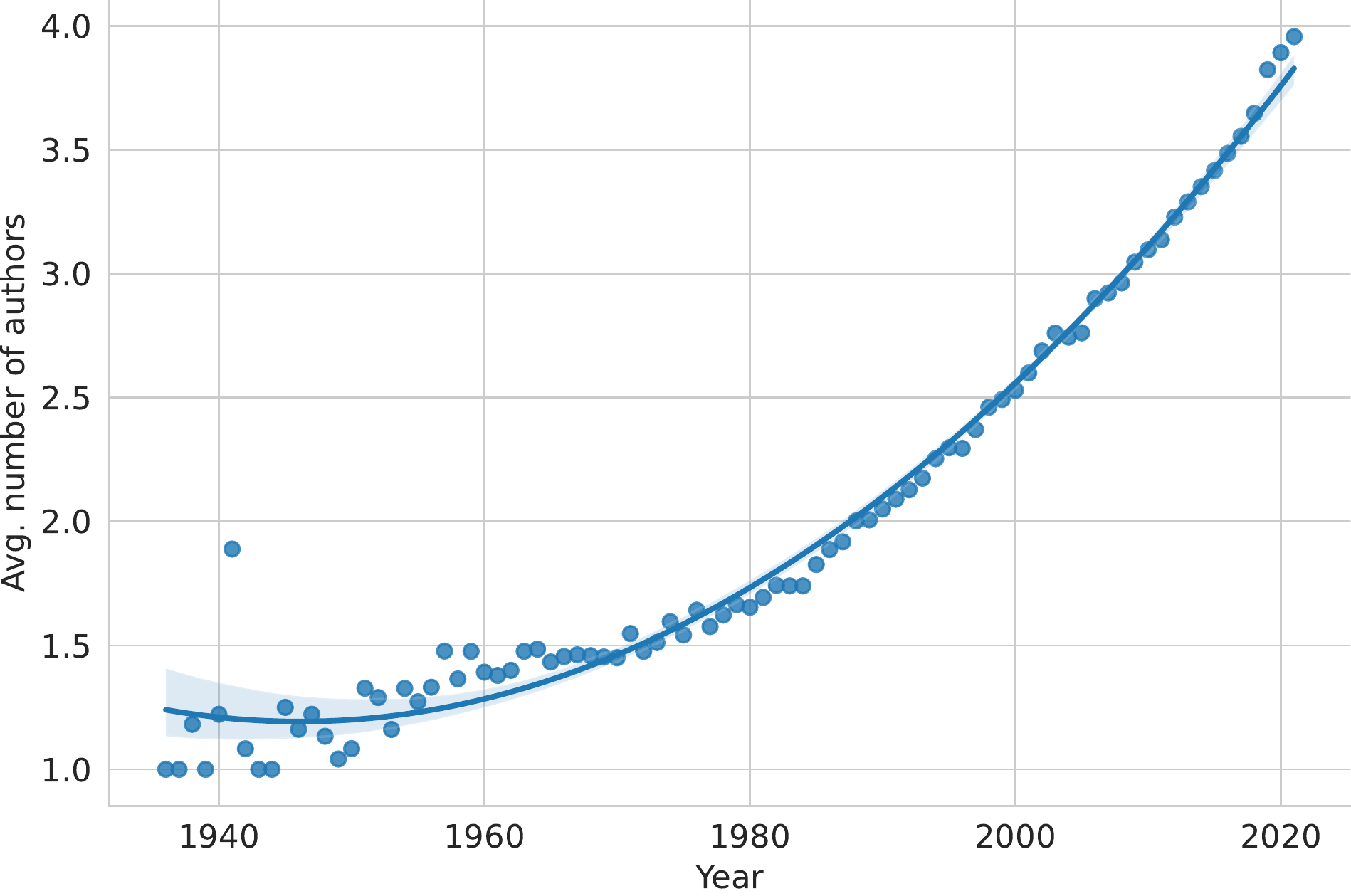}
    \caption{A cubic approximation on the average number of authors per paper between 1936 and 2021.}
    \label{fig:analysis-num-authors-per-paper}
\end{figure}

Looking at the average number of authors on a paper over time (\Cref{fig:analysis-num-authors-per-paper}), we observe a steep increase from 1970 onward.
The increase in authors on the same paper indicates growing collaboration in computer science; a positive sign for a healthy and growing research area.\\

\begin{figure}
    \centering
    \includegraphics[width=\columnwidth]{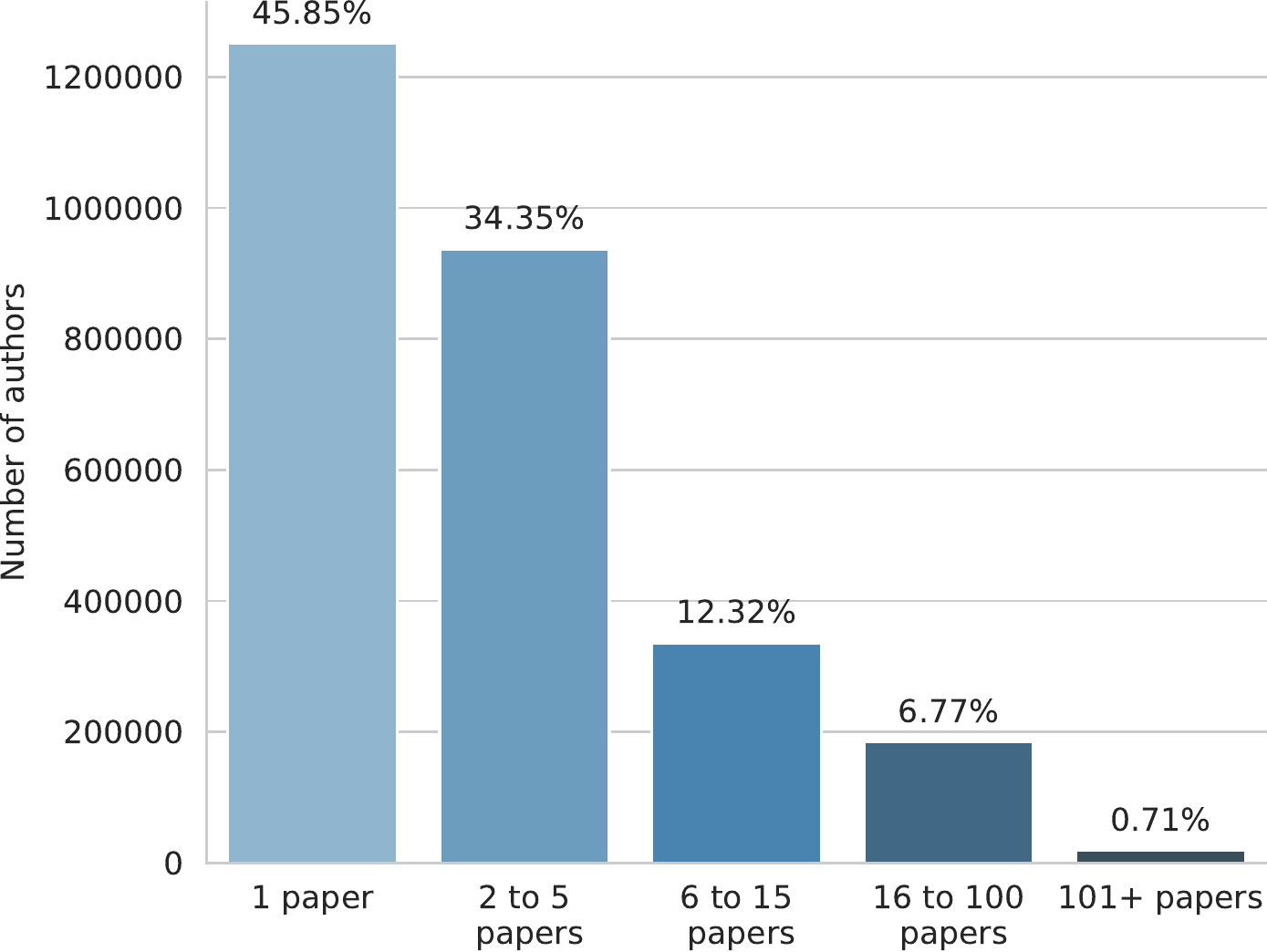}
    \caption{The number of authors and their published papers until December 2021.}
    \label{fig:analysis-publications-by-authors}
\end{figure}

\textit{Q3. How many authors published one or more papers?}\\[1.5pt]
A. More than 1.4 million authors have published precisely one paper in DBLP.
\Cref{fig:analysis-publications-by-authors} shows the number of \projectName authors corresponding to different number-of-papers bins. 
Similar bins were used in the analysis of NLP papers in \cite{Mohammad20b}.

Analysis of both \projectName and \aclAnthology \cite{Mohammad20b} shows that most authors publish exactly one paper. 
Though, the skew towards single-publication authors is more stark in the \aclAnthology (57.9\%) compared to \projectName (45.85\%).
The NLP field has seen a substantial surge in new authors recently, and we hypothesize their gain is higher than the computer science average, explaining the greater skew in \aclAnthology.
If we consider the sum of authors who published two or more papers, \projectName shows 54.14\%, while \aclAnthology only shows 42.10\%, corroborating our assumption.
The same behavior in the number of authors is also observed when using the same bin split as in \cite{Mohammad20b}.\\

\begin{figure}
    \centering
    \includegraphics[width=\columnwidth]{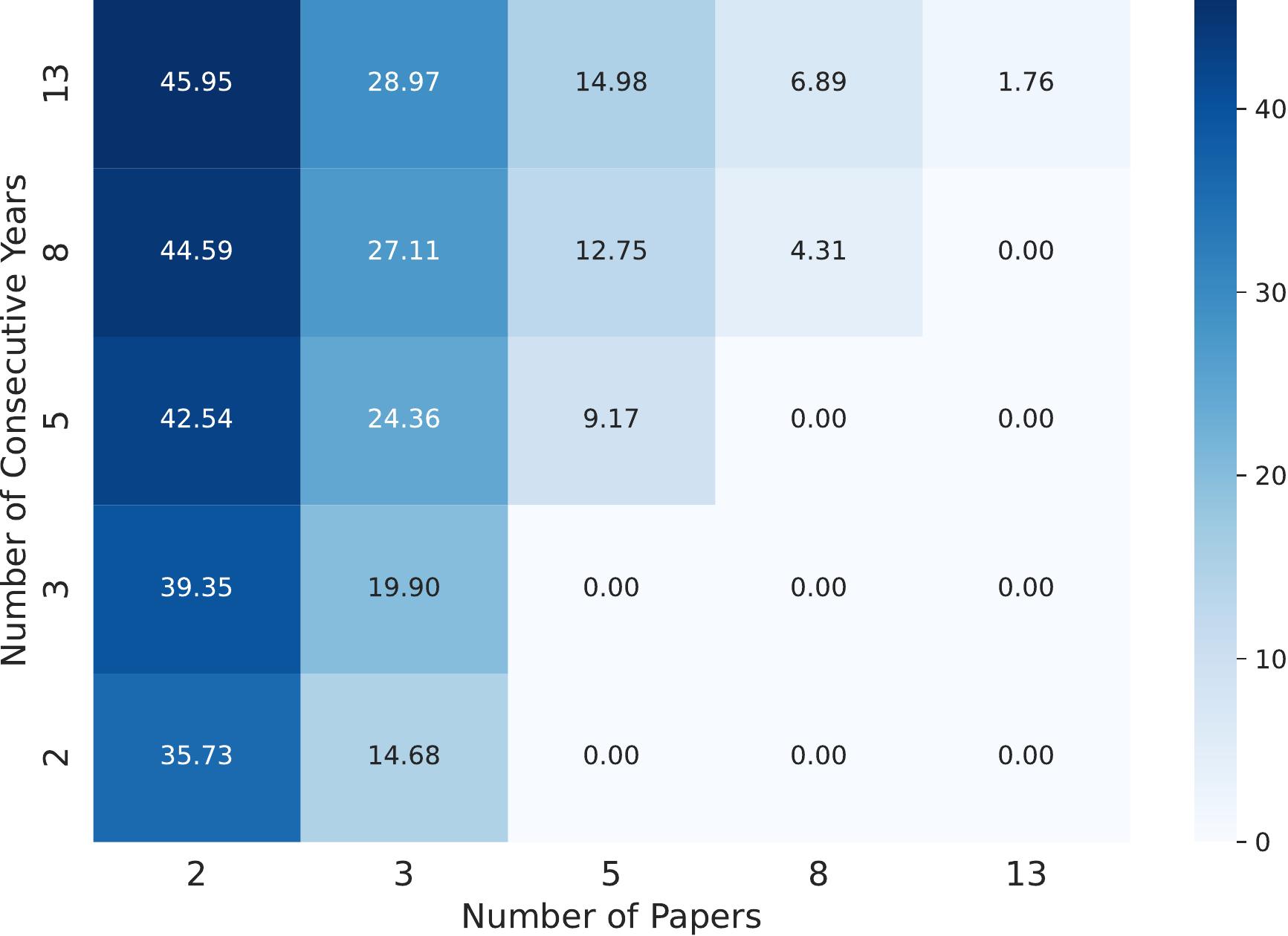}
    \caption{The relative amount of active researchers (colored and in \%). An active researcher is defined by the minimum number of papers published (x-axis) in a number consecutive years (y-axis). For example, in the last 13 years, out of all researchers who published in that time, 45.95\% published 2 or more papers.}
    \label{fig:analysis-active-researchers}
\end{figure}

\textit{Q4. Who is actively publishing in DBLP?}\\[1.5pt]
A. To answer this question, we identify authors that published at least a certain amount of papers over the last years starting from 2021.
In our initial investigation, we measure the number of authors that published at least x=\{2, 3, 5, 8, 13\} papers in the last y=\{2, 3, 5, 8, 13\} consecutive years (before 2021).
Intuitively, the more papers are published by an individual author in a specific time range, the more active this researcher is.
\Cref{fig:analysis-active-researchers} shows the results visualized in a heatmap.
We find the highest proportion of active researchers with 13 consecutive years and 2 or more papers published (45.95\% of authors).
The results seem more sensitive to changes in the number of published papers ($x$-axis) than the time range ($y$-axis).
When increasing the number of papers to 3 or more, the number of active researchers drops to 28.97\%, while when decreasing the time to 8 years, it remains at 44.59\%.
We assume the typical time in which researchers publish actively is relatively short as a significant proportion of research is performed with a limited time horizon (e.g., Ph.D. students).
The increase in the number of active researchers gets smaller as we increase the number of consecutive years considered.
If we assume a doctorate degree takes, on average, between 5 and 8 years\footnote{This time can vary depending on the program and area.} the decline in the difference between years (especially from 8 to 13) for active researchers can be justified.
In the last 13 years, less than 2 out of 100 researchers (1.36\%) published 13 or more papers, showing that few authors remain being active in academia for more than a decade.
This experiment shows many other possibilities of \projectName, such as investigating how many of the researchers move from the first author to other positions, indicating their role in the research has changed. 

\begin{figure*}
    \centering
    \begin{subfigure}{.5\textwidth}
        \centering
        \includegraphics[width=.92\textwidth]{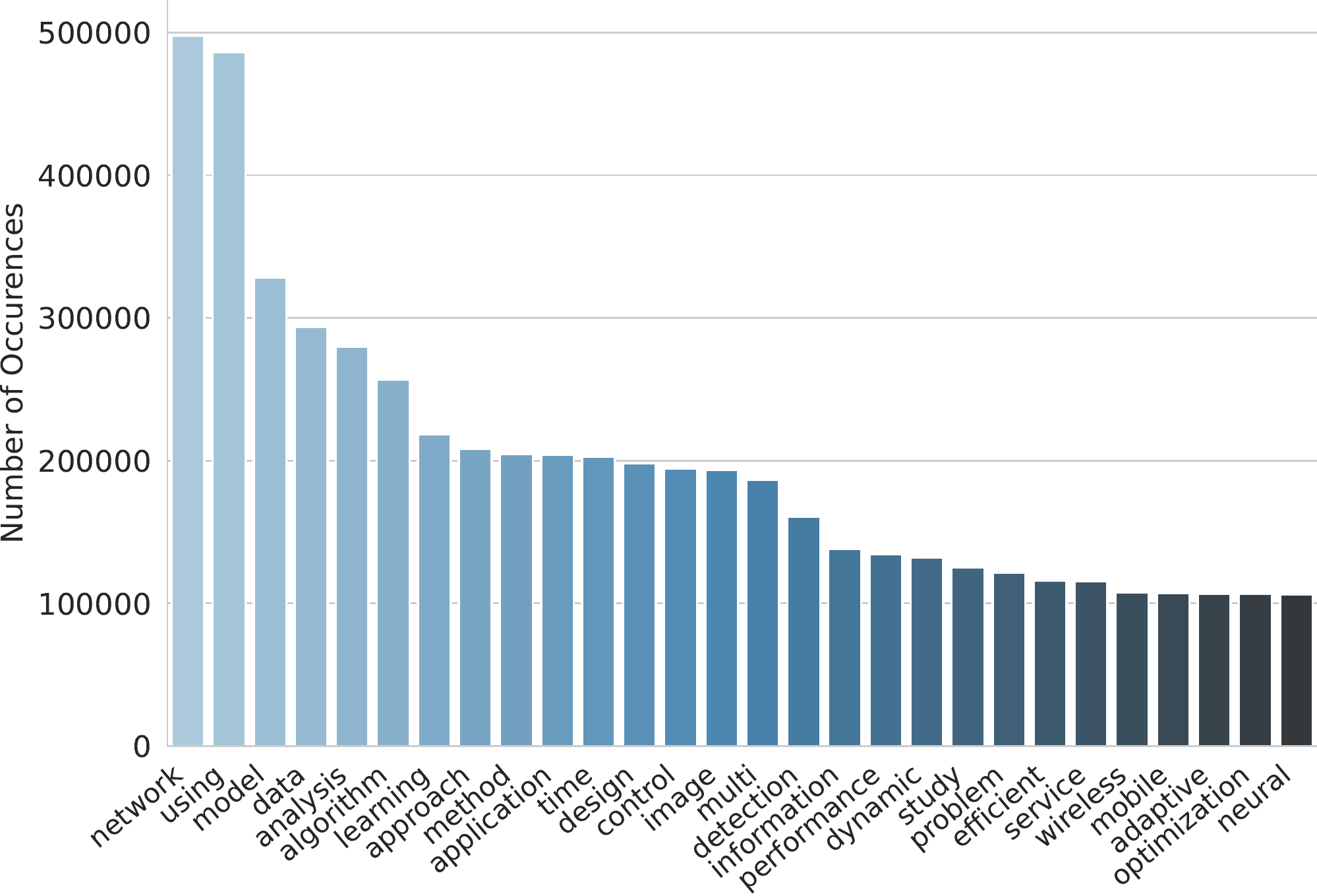}
        \caption{Frequencies of terms in titles.}
    \end{subfigure}%
    \begin{subfigure}{.5\textwidth}
        \centering
        \includegraphics[width=.92\textwidth]{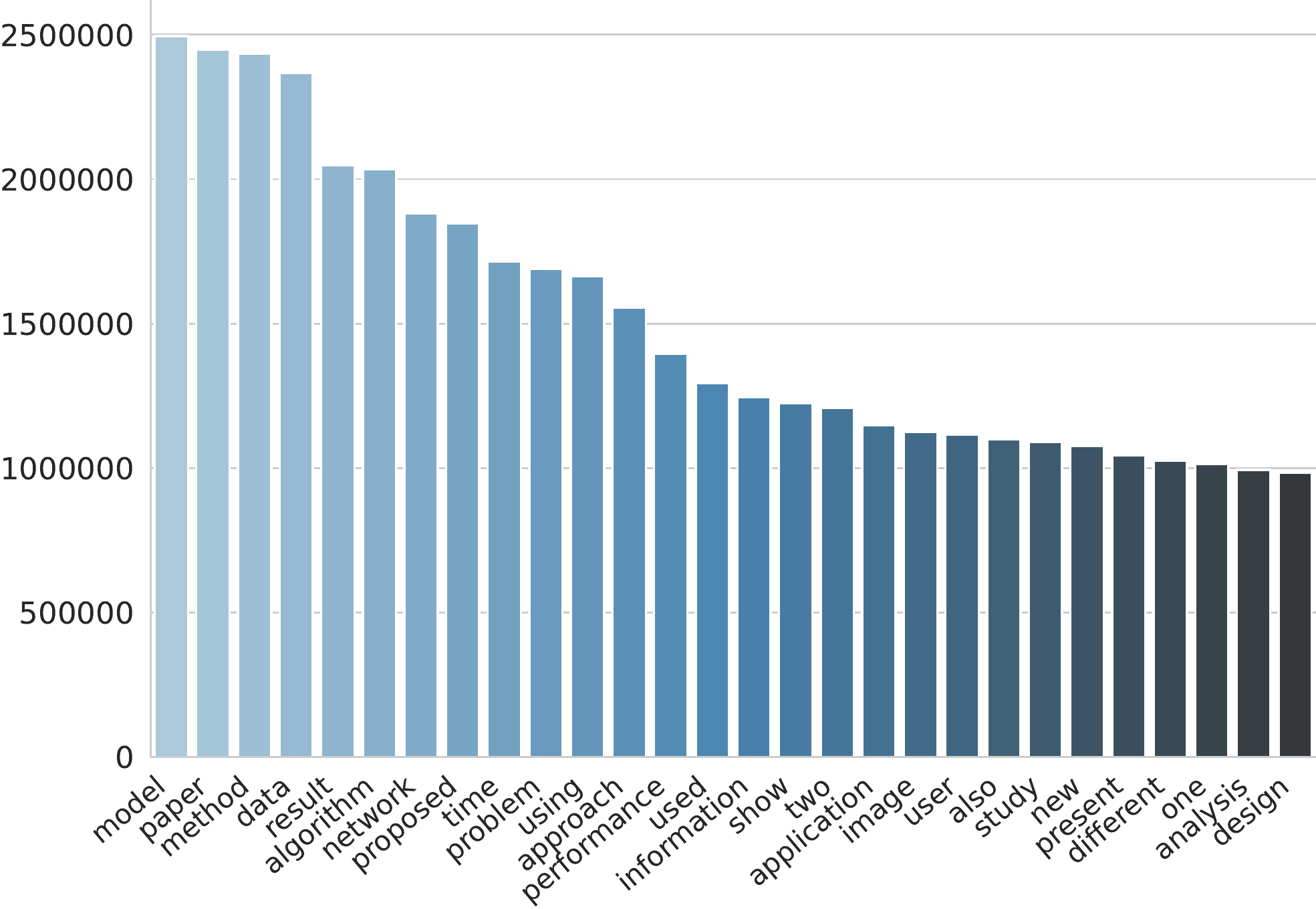}
        \caption{Frequencies of terms in abstracts.}
    \end{subfigure}%
    \caption{The most common terms in titles and abstracts between 1936 and 2021.}
    \label{fig:analysis-term-frequencies}
\end{figure*}

\begin{figure*}[htb]
    \center
    \includegraphics[width=.92\textwidth]{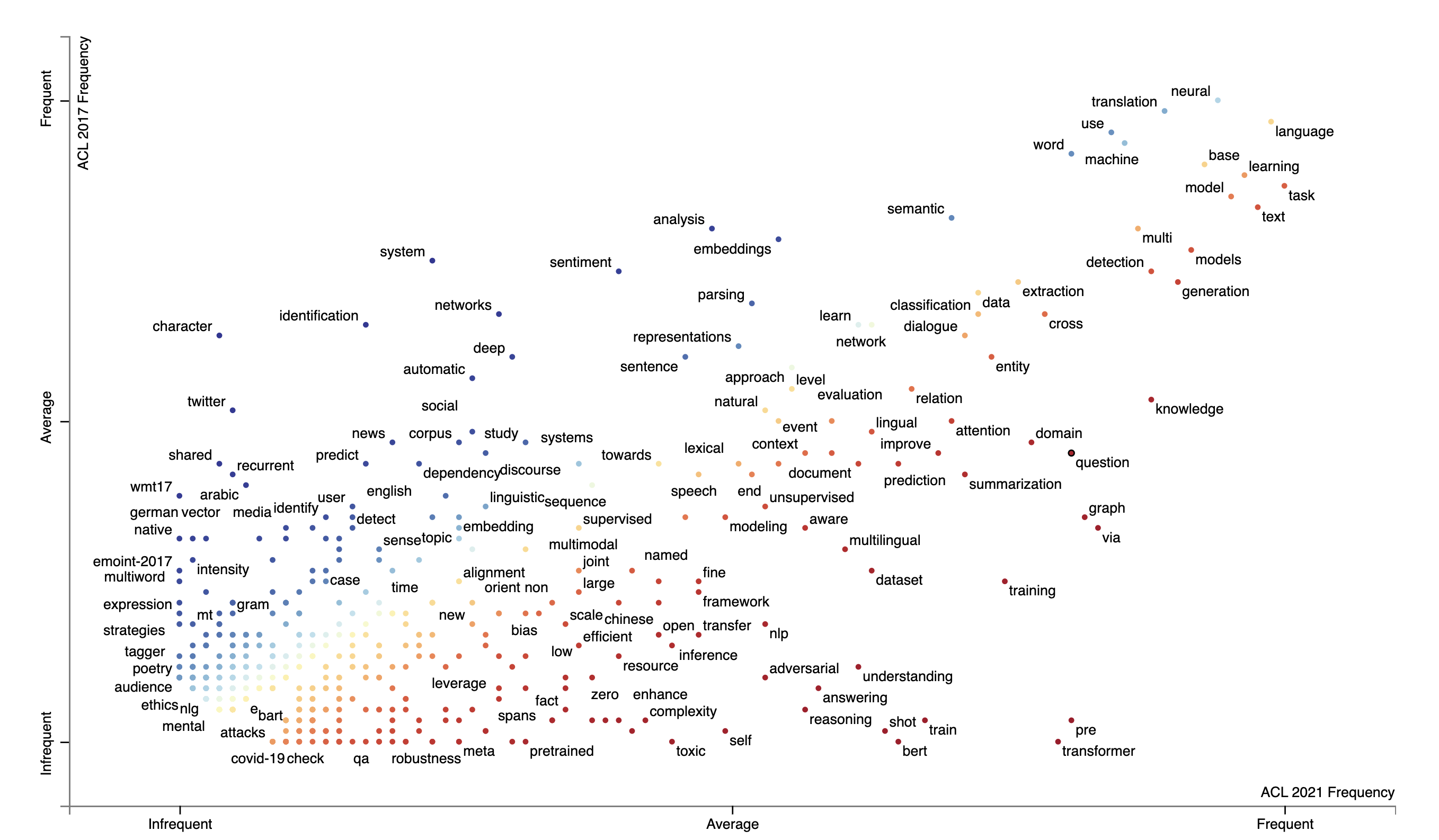}
    \caption{The most common unigrams in abstracts of ACL between 2017 and 2021.}
    \label{fig:analysis-acl-scatterplot}
\end{figure*}

\subsection{Trends in Topics}


\textit{Q5. Which are the most common terms in titles and abstracts, and how do they differ?}\\[1.5pt]
A. We visualize the 30 most frequent unigram lemmas (without stopwords) of titles (left) and abstracts (right) in \Cref{fig:analysis-term-frequencies}.
Our first observation is that the frequent words in titles convey key research topics and are not filler words or discourse connectives.
Abstracts contain filler words such as ``also'', ``present'', ``new''.
The 5 most frequent words in both abstracts and titles contain ``model'' and ``data'', indicating an increase in importance for data to obtain effective models in computer science applications.
The term ``network'' is represented frequently in abstracts and titles, which might be related to both network analysis and neural networks.
Our findings suggest the rising of machine learning if we consider terms such as ``learning'', ``optimization'', and ``neural''.
A key difference between abstracts and titles is that the former contains additional information about the paper content, reflected in lemmas such as ``performance'', ``time'', and ``results''.
In the future, we plan to investigate the semantic representation of terms in titles and abstracts to understand and compare their content.\\

\begin{figure*}
    \centering
    \begin{subfigure}{.5\textwidth}
        \centering
        \includegraphics[width=.95\textwidth]{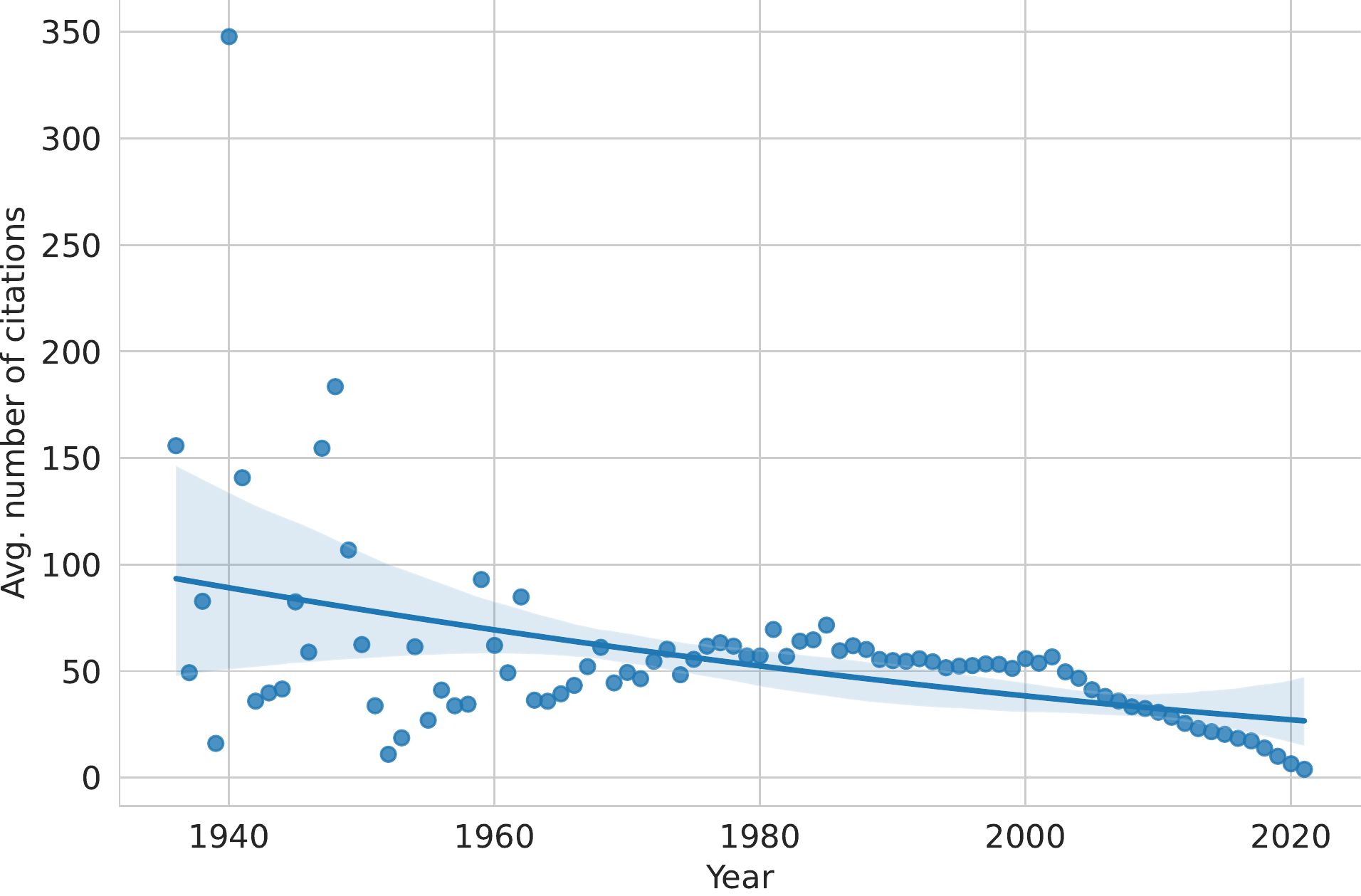}
        \caption{The average number of incoming citations.}
    \end{subfigure}%
    \begin{subfigure}{.5\textwidth}
        \centering
        \includegraphics[width=.95\textwidth]{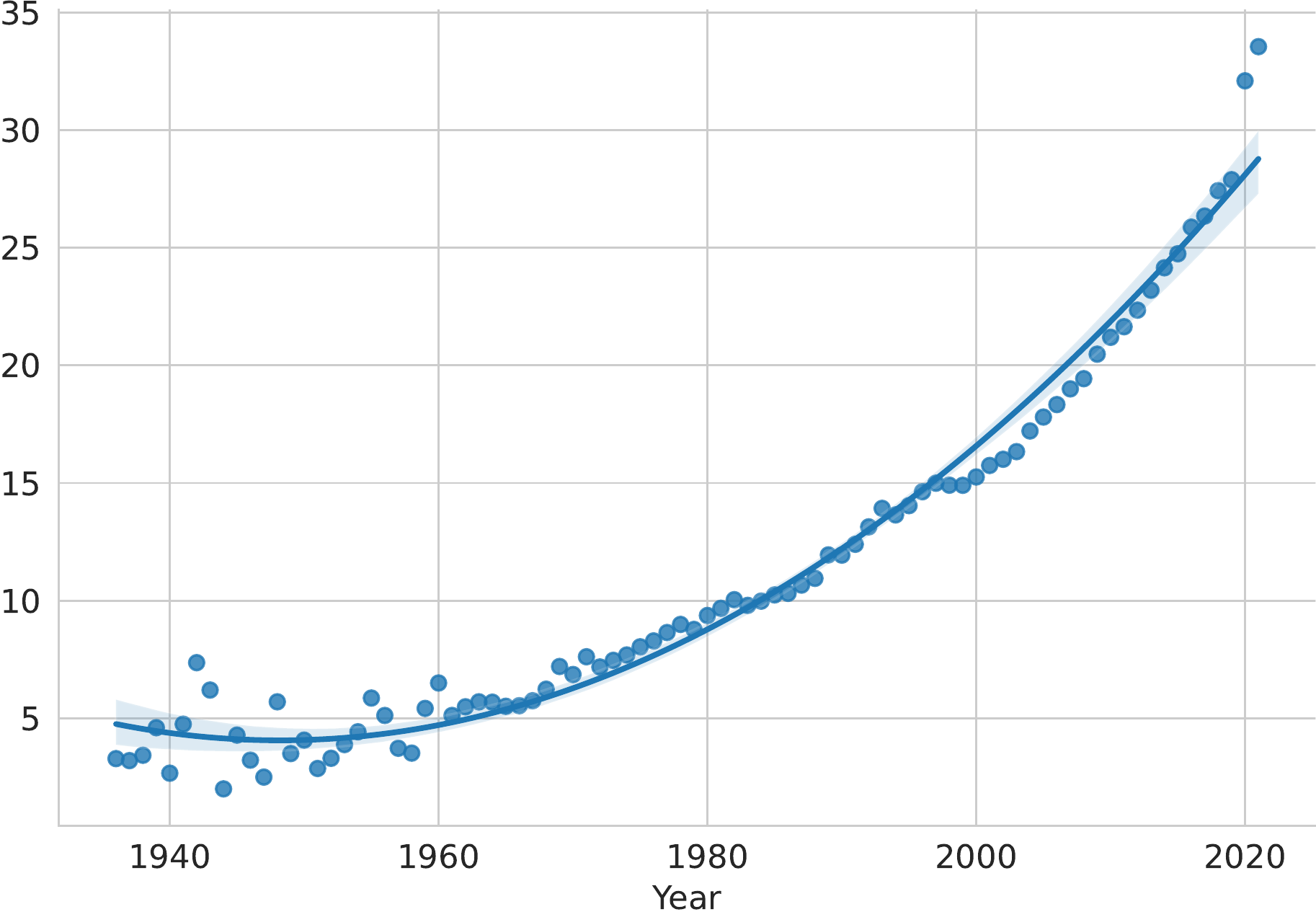}
        \caption{The average number of outgoing citations.}
    \end{subfigure}%
    \caption{A cubic approximation on the citations of papers published between 1936 and 2021.}
    \label{fig:analysis-citations-by-years}
\end{figure*}

\textit{Q6. Which topic trends are described in abstracts?}\\[1.5pt]
A. In \Cref{fig:analysis-acl-scatterplot}, we show the occurrence of unigram term frequencies of abstracts for ACL from 2017\footnote{2017 was chosen as a starting point because of the publication of the highly influential Transformer model in \cite{NIPS2017_3f5ee243}.} (y-axis) and 2021 (x-axis).
We chose ACL as it is one of the largest and most influential conferences in NLP, a major sub-field of computer science.
Terms close to the diagonal represent similar frequencies among both years.

We find ``summarization'', ``dialogue'', and ``topic'' close to the diagonal as their studies are prevalent in 2017 and 2021.
Data points located on the most right of the x-axis and at the same time on the lower part of the y-axis indicate an emerging topic, whereas the converse suggests that the interest in a topic is declining.
For example, in the lower right, terms such as ``bert'', ``transformer'', and other related prefixes (e.g., ``pre'' for ``pre-training'', ``fine'' for ``fine-tuning'', or ``masked'' for ``masked language modeling'') frequently appear in 2021 but rarely appear in 2017. While these terms were not popular in 2017, four years later, 90 papers mentioned ``bert'' and 64 papers mentioned ``transformers'' at least once in their abstracts.
This finding is in line with a recent trend of the Transformer model published in 2017 \cite{NIPS2017_3f5ee243}, and particularly the popularity of BERT which was published in 2018 (arXiv) / 2019 (NAACL) \cite{devlin-etal-2019-bert}.
Assuming Transformer-based models will continue to be applied in 2022, a new version of \Cref{fig:analysis-acl-scatterplot} would show ``transformer" even further on the x-axis if compared to 2017, or closer to the diagonal if 2021 was considered.
Another group of terms is related to ``covid-19'', the viral disease of SARS-CoV-2 first measured in late 2019 and affecting countries worldwide since early 2020\footnote{\url{https://bit.ly/3GrWojh}}.

Terms that reduce in frequency compared to the previous four years are for example ``sequence'', ``lstm'', ``recurrent'', ``vector'', and ``embedding'' which can be related to traditional recurrent sequence models and static word embeddings used before dynamic models (e.g., Transformer).
Also, terms such as ``parsing'', ``dependency'', or ``convolutional'' reflect a decrease in interest in of dependency parsing and convolutional neural networks in NLP applications.

Analyzing term frequencies is the first step to understanding topics and trends in \projectName.
In the future, we will combine term frequencies with topic models to better understand computer science papers' contents and the relation between their topics.

\subsection{Citation Patterns} \label{sec:analysis-citations}
\begin{figure}
    \centering
    \includegraphics[width=\columnwidth]{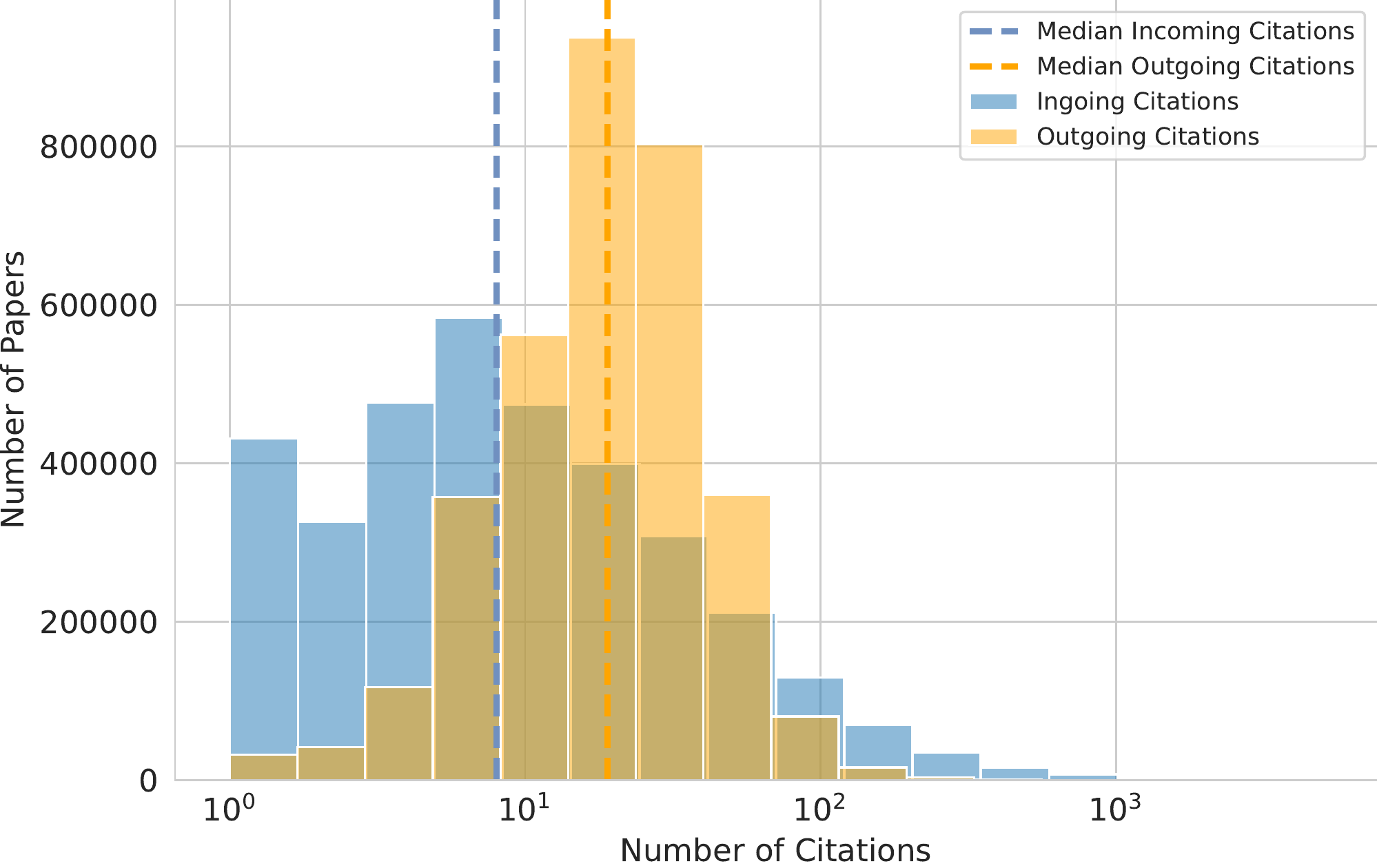}
    \caption{The distribution of incoming and outgoing citations published between 1936 and 2021.}
    \label{fig:citation_distribution}
\end{figure}
\textit{Q7. How many sources do papers use, and how many citations do papers acquire?}\\[1.5pt]
A. \Cref{fig:citation_distribution} shows the distribution of incoming citations (i.e., how often a single paper was cited) and outgoing citations (i.e., the number of bibliography entries in a single paper) for all papers in \projectName.
The average number of incoming and outgoing citations are 28.16 and 22.95, respectively.
Even though the average citation count for both incoming and outgoing are similar, their distribution is quite different.
While outgoing citations are seemingly normal distributed, incoming citations are skewed to the left.
When considering the median of the incoming citations (8), we observe that most papers achieve less than a hundred citations.
Only a few papers reach the thousand mark.
A fair proportion of papers receive no citations (1,921,844, or 30.06\%) or only 1 citation (431,227, or 6.75\%).
For the outgoing citations, the median (18) is closer to the average and few papers' bibliographies contain more than a hundred entries.
These results are valid for papers citing others within \projectName. To measure how many incoming citations are from papers of other fields outside of \projectName, we use the \semanticScholar API with the result that 21.15\% of citations come from papers outside of \projectName.\\

\textit{Q8. Are we citing more papers than in earlier decades? Do recent papers receive more citations than papers published earlier in the past?}\\[1.5pt]
A. Tracking citations over time reveals two patterns.
\Cref{fig:analysis-citations-by-years} shows the trend of average incoming (left) and outgoing (right) citations over the time between 1936 and 2021.
We observe that the period before mid 1960s has much more variability in terms of incoming citations, likely because of a much smaller number of papers from that period.
After the mid 1960s, we see a rather steady average incoming citations in the low 50s, and a sharper decline for papers since the early 2000s.
The trend for papers published in 2000s is likely because the more recent papers have had less time to gain popularity and accumulate citations.
With time, we expect the average to go markedly more up for the papers published in the 2000s (as opposed the pre-2000 papers).
When looking at the number of citation entries in a paper, the trend is different, as publications have increasingly used more references in their works over the years.
With an increase of proceedings publishing primarily online, publishers have no additional cost to include extra pages, and therefore progressively introduce citation-friendly rules such as additional or unlimited pages for references.
Another factor for the steep increase in outgoing citations results from the growth of computer science as a field over time, and therefore the usage of earlier research.

\section{Further Applications}
\projectName has numerous applications.
In the following, we describe the most interesting ones for future work.\\[1.5pt]
\textbf{Topic Analysis.} Identifying topics of publications and tracking their distribution over time for venues, authors, and affiliations enables insights into their popularity.
Schumann \shortcite{schumann-2016-brave} provides initial attempts in modeling term life cycles and clustering terms using the \aclAnthology Reference Corpus \cite{radev-etal-2009-acl}.
Further questions could target the identification of trendsetters (e.g., innovative and influential authors) and their respective followers.
A trendsetter could be a venue offering a particular topic in their call for papers or an author publishing on an emerging topic.
We are also interested in understanding whether venues or authors follow each other's topics over time and the possible reason behind this.\\[1.5pt]
\textbf{Influence of research fields.} We showed that most citations in \projectName come from computer science publications.
However, the influence of computer science on other fields (e.g., medicine, psychology) is still unclear.
Papers may focus on advancing state-of-the-art in a computer science method (e.g., object detection) but with a focus in another field (e.g., cancer detection from x-ray images).
By understanding the papers' content and its citations, we can estimate the relative influence of research fields on a paper.
The previous example paper on object detection could mainly cite computer science papers, but their contents may strongly connect to the medical domain.
Computing correlations and influences of papers will make interdisciplinary research more transparent, allowing us to understand its trends and collaborations.\\[1.5pt]
\textbf{Impact, success, and productivity.} In this work, we analyzed citations in an exploratory manner to measure their impact over the years.
The question of how we can define successful authors, venues, and affiliations is yet to be answered.
A compelling direction of future work is to identify the influence of various features (above and beyond citations) to arrive at a more robust picture of influence of a scholar or a field of study.
Also, we categorized active researchers according to their publication count over a period of time.
To estimate their productivity in the future, we could identify their output and relate it to their career span defined by the first and last year of publication.\\[1.5pt]
\textbf{Gender gap and fairness.} An increasing number of problems for researchers and society have their roots in ethical issues.
Previous studies investigated the gender gap within science and its publications.
Mohammad \shortcite{Mohammad20c,Mohammad20d} found only 29\% of first and 25\% of last authors are female.
Other questions that \projectName enables to answer are fairness about locations and ethnicity.
Is there a bias for accepted papers for researchers from wealthy countries?
Are publications cited less because they are not originating from prestigious universities and companies?
In the future, we also want to extend our dataset with openly reviewed papers to understand the acceptance rate considering fairness criteria.

\section{Conclusion}
We created a new resource, the DBLP Discovery Dataset (\projectName), that contains metadata associated with over 6.3 million computer science papers.
We also conducted experiments to explore a number of questions on the broad trends of computer science publications.
Notably, we showed that while computer science is enjoying a growing popularity and attracts increasingly more authors, the proportion of researchers remaining active in the field for a long time is rather small.
We demonstrated, that the number of citations a paper receives declines in the last decades while papers include more sources in their bibliographies.
Furthermore, the distribution of citations shows that most papers receive few citations (less than 10 and often none), while few papers reach more than a thousand citations.
By analyzing the most common terms in abstracts and titles, we showed that titles convey key research topics and abstracts revealed recent topic trends of NLP, such as an increased usage of the popular Transformer.

In the future, we want to provide \projectName through a REST API to answer specific queries (e.g., retrieving papers with more than, say, 10 citations by authors of a user-chosen affiliation).
To make access to our dataset intuitive and without specific hardware requirements, we also want to release an interactive web tool.
At the time of writing, we work on a topic analysis micro-service that generates topic distributions of venues, authors, and individual publications using generic model configurations.
The findings, datasets, and source code will always be publicly available for research purposes.

\section{Acknowledgments}
This work would not be possible without the great resources offered by DBLP, \aclAnthology, \semanticScholar and teams, to whom we are very thankful. We also thank Lennart K\"{u}ll who helped us during the initial phase of the project. 

\section{Bibliographical References}
\bibliographystyle{lrec2022-bib}
\bibliography{main}

\section*{Appendix}
\renewcommand{\thesubsection}{\Alph{subsection}}

\subsection{Ethical Concerns \& Broader Impact}

As we explore authors and their scientific publications in computer science, \projectName's data (e.g., author's names, affiliations, web pages) are not anonymized.
We cannot include abstracts from restricted access papers for the publicly available version as they fall under the same copyright as full-texts.
All remaining material available in \projectName is licensed to the general public under a copyright policy that allows unlimited reproduction, distribution and hosting on any website or medium\footnote{\url{https://dblp.org/db/about/copyright}}.
Currently our experiments and findings hold truth to publications inside of DBLP which is a expressive subset of computer science publications.
However, some publications are not indexed by our dataset, and therefore do not provide a complete picture of the computer science field.
We advise the use of \projectName with carefulness and attention, as it contains sensitive information from real people.

We believe our approach can be transferred to any domain where its data is organized and available. 
Therefore, we hope other major publishers (e.g. Elsevier) acknowledge and adopt the benefits of open policies with respect to their repositories. 
Open access to other publishers' data would unveil new possibilities to our investigations. 
We see medicine and education as areas with great potential to apply our research. 
The COVID-19 pandemic has shown the benefit of publicly accessible information, as new discoveries were released every day; and thereby increasing the collective understanding about the topic and supporting the creation of solutions (e.g., vaccines, treatments, prevention measures). 
Other infirmities (e.g., dengue-fever, AIDS, cancer) could also take advantage of such collaborative efforts.

\end{document}